\documentclass[11pt,twoside]{article}
\usepackage{asp2021}
\usepackage{hyperref}

\aspSuppressVolSlug
\resetcounters

\begin{document}

\title{Different Stories: European Astronomy as a Tool of Colonization in the Americas and Hawai’i}
\author{Gregg A. Wade$^{1,2}$, Danielle Lussier$^2$
\\
\affil{$^{1}$Royal Military College of Canada, Kingston, Ontario, Canada; \email{wade-g@rmc.ca}}
\affil{$^2$Queen's University, Kingston, Ontario, Canada; \email{danielle.lussier@queensu.ca}}}

% This section is for ADS Processing.  There must be one line per author.
%\paperauthor{Sample~Author1}{Author1Email@email.edu}{ORCID_Or_Blank}{Author1 Institution}{Author1 Department}{City}{State/Province}{Postal Code}{Country}
%\paperauthor{Sample~Author2}{Author2Email@email.edu}{ORCID_Or_Blank}{Author2 Institution}{Author2 Department}{City}{State/Province}{Postal Code}{Country}
%\paperauthor{Sample~Author3}{Author3Email@email.edu}{ORCID_Or_Blank}{Author3 Institution}{Author3 Department}{City}{State/Province}{Postal Code}{Country}

\begin{abstract}
The history of Western astronomy is frequently told as a story of discovery, technological innovation, and human progress. Yet the astronomical knowledge, instruments, and techniques associated with these achievements also contributed to European colonial expansion. Through four material and visual examples - a marine chronometer, an astronomical surveying quadrant, the \emph{Nautical Almanac}, and contemporary observatories on Maunakea - we invite reconsideration of astronomy's relationships with navigation, territorial mapping, colonial authority, and Indigenous lands. How might these histories be told differently when considered through Indigenous Knowledges and Perspectives? What might such reconsideration reveal about the continuing relationships between astronomy, colonialism, and place?
\end{abstract}

\section{Different stories}

The stories we tell are important. The history of Western science is often presented as a triumph of light over darkness, in which science and scientists are situated as noble figures working in the service of progress. Consider the familiar accounts of Copernicus and Galileo standing fast against the Catholic church in their support of the Heliocentric model of the solar system; of Giordano Bruno’s execution for espousing the principles of Cosmic Pluralism; of Isaac Newton, his apple, and the Law of Universal Gravitation; or, the tragic tale of Marie Curie and her discovery of radioactivity. Such stories foreground intellectual courage, ingenuity, and discovery. They can also encourage us to imagine science as an enterprise separate from the political and social conditions in which it develops.

But storying astronomy's ideas and instruments during the European so-called ``Age of Discovery" and its aftermath - including the invasion and colonization of territories now known as the United States - tells a different tale. Astronomical knowledge was not merely carried along on voyages of exploration. It helped make those voyages possible, made distant territories more readily measurable and representable, and contributed to the infrastructures through which colonial powers asserted authority over lands and peoples. These relationships deserve particular attention in the context of the 250th anniversary of the founding of the United States, an event that can be understood quite differently when viewed alongside the historic and contemporary lived experiences of Indigenous peoples.

This short paper invites the viewer to consider four images (Figs. 1-4) and to imagine how the stories prompted by them might differ, in fundamental ways, from the perspectives of scientists and colonized peoples. We do not suggest that the objects possess a single meaning, or that a single narrative can represent diverse Indigenous perspectives. Rather, we engage in this initial reflection to pose questions about astronomy's material, institutional, and political histories.

\section{Instruments, measurement, and colonial power}

The first object is the marine chronometer K1, carried aboard James Cook's \emph{Resolution} during his second Pacific voyage (1772-1775). Accurate timekeeping, together with celestial observations, provided a means to determine longitude at sea and thereby improved the reliability of navigation and chart-making \citep{Sorrenson1996}. Cook reportedly called K1 ``our trusty friend'' and ``never-failing guide'' \citep{RMG_K1}. In a conventional history of science, the chronometer exemplifies the extraordinary achievements of precision instrument making in 16th-18th century Europe. Another perspective asks what the capacity to navigate, return to, and map distant places meant for the peoples already living there. Scientific achievement and colonial consequence were not mutually exclusive: the same instrument can be celebrated for its accuracy and examined for its role within imperial maritime expansion.

The second object is a quadrant attributed to Peter and John Dollond Co. and associated with surveying in Lower Canada between approximately 1780 and 1820 \citep{SMG_Quadrant}. Astronomical and geodetic observations supplied positions and bearings that could be incorporated into surveys and maps. Such practices made vast territories comprehensible and measureable, hence subject to framing within European systems of property, administration, and territorial jurisdiction \citep{Edney1997}. While the act of measurement did not itself establish legitimate ownership, mapping and surveying could be mobilized within colonial processes to displace Indigenous relationships with lands and waters and replace them with imposed boundaries. The instrument therefore prompts questions not only about \emph{how} land was measured, but also \emph{who} authorized the measurement, \emph{whose} understandings of place were recognized, and \emph{to what ends} the resulting representations were put.

The third object is the title page of the {\it Nautical Almanac} for 1767, published by order of the Commissioners of Longitude \citep{NauticalAlmanac1767}. The almanac made predictions of celestial positions and other astronomical information available in standardized tabular form, notably supporting the lunar-distance method for determining longitude \citep{Sadler1976}. Its significance lay not in a single instrument or observer but in a system of observations, developments of physical theory, mathematical calculations, publishing, and maritime practice. The production and circulation of such knowledge helped translate astronomical science into repeatable techniques useful to navigators and surveyors. It reminds us that colonial power depended not only on individual explorers, but also on institutions and systems of expertise that supported their activities.

Together, these three objects illustrate distinct but connected aspects of the application of astronomy to colonial projects: navigating across oceans, locating and mapping territories, and institutionalizing the knowledge needed to do both. Their histories should not be reduced to a claim that astronomical science inevitably produced colonization. Rather, they demonstrate how scientific capabilities were developed, funded, and deployed within specific imperial circumstances, with consequences that were experienced in fundamentally different ways by different peoples.

\section{Space, place, and scientific authority}

The fourth image brings the conversation into the present. It shows the Subaru and Keck observatories on Maunakea, Hawai\textquoteleft i, in 2019, with a stone marker in the foreground. Maunakea is a place of profound cultural and spiritual significance for many Native Hawaiians, as well as a site of internationally important astronomical research. Disputes surrounding telescope development and stewardship have raised questions about consent, governance, cultural responsibilities, and the authority to determine how lands should be used \citep{CasumbalSalazar2017,TanigawaLau2022}. This image invites us to examine a recurring assumption: that the scientific value of a place can be defined independently of the people whose histories, responsibilities, and relationships are rooted there. The image asks us to look beyond the telescopes as technical achievements and to consider the mountain as a place with meanings that cannot be captured by astronomical site assessments or publication counts.

Attention to Indigenous Knowledges and Perspectives can challenge the apparent neutrality of scientific representations of space. A chart, a survey, a coordinate, and a telescope site each make certain relationships visible while potentially obscuring others. Indigenous Research Methodologies emphasize the relational character of knowledge and the responsibilities that accompany its production and use \citep{Kameeleihiwa1992,Smith2012,Wilson2008,ForsytheMarkides2026}. Asking what stories like those briefly considered here leave out is not a rejection of astronomical knowledge. It is an invitation to understand its situated uses and consequences, and to recognize that other knowledge traditions and relationships to place possess their own authority. 

\section{Reconsidering the stories we tell}

We began with a question: how were the instruments and techniques of astronomy implicated in the production of colonial knowledge, authority, and territorial control in the Americas and the Pacific? The objects presented here do not provide a complete answer. Instead, they offer points of entry into a broader inquiry at the intersection of the history of science, material culture, colonial history, and Indigenous Knowledges.

The familiar story of a chronometer, quadrant, or almanac emphasizes precision, ingenuity, and the expansion of geographical knowledge. Different stories may foreground the invasion of Indigenous territories, the displacement of existing systems of knowledge and governance, and the use of measurement to support colonial claims. Similarly, the story of an astronomical observatory can be one of scientific discovery while also being a story about Indigenous sovereignty and responsibilities to place.

Reconsidering these histories requires more than introducing different perspectives into established narratives of scientific progress. It also invites reflection on the relationships and responsibilities through which knowledge is produced, interpreted, and shared. \citet{Wilson2008}'s conception of relational accountability offers one framework for approaching this work.

\clearpage
\begin{figure}[!h]
\centering
\begin{minipage}[t]{0.48\textwidth}\centering
\includegraphics[height=2.45in,keepaspectratio]{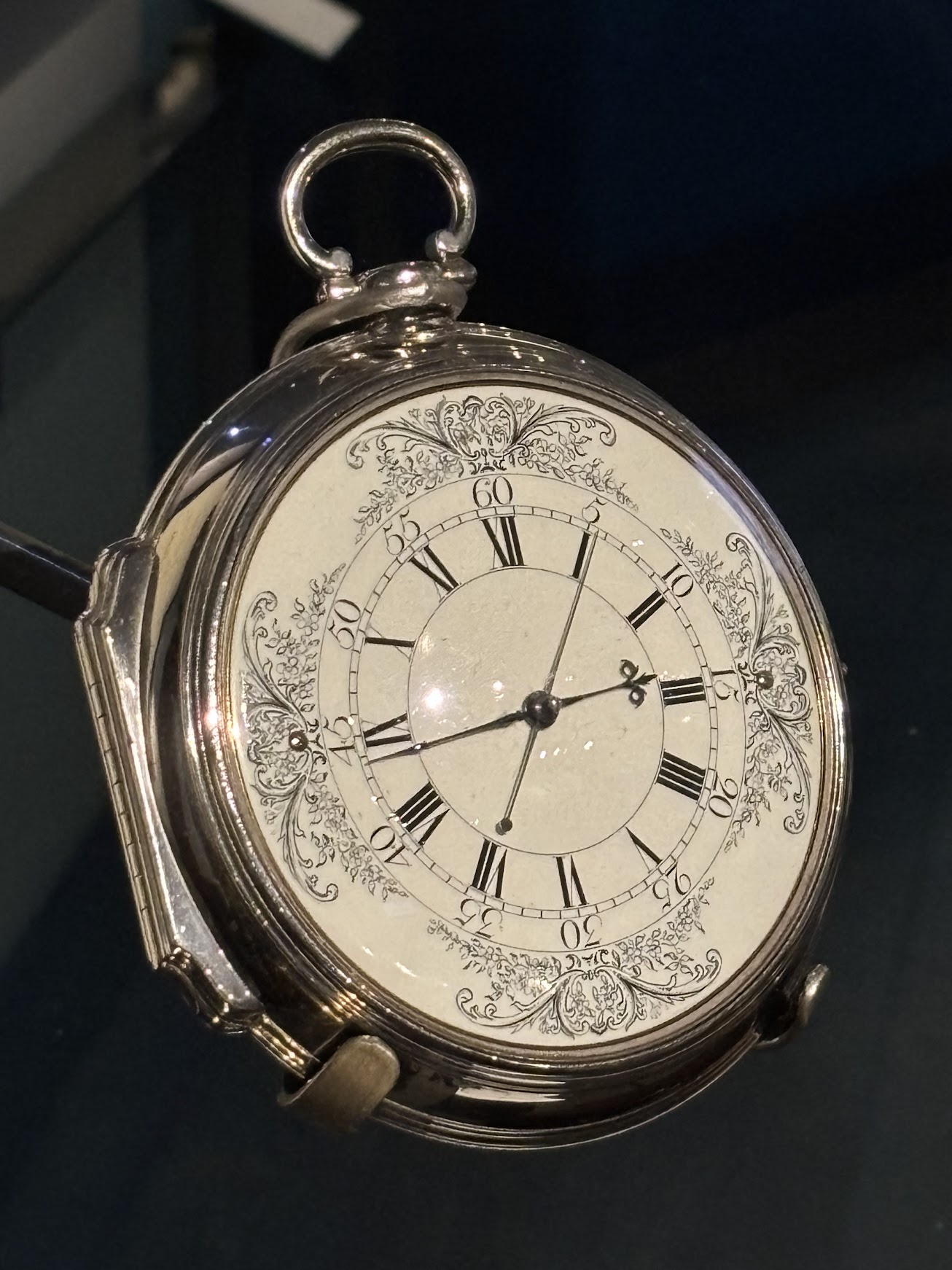}
\caption{\textbf{(a)} Marine chronometer K1, carried aboard Cook's \emph{Resolution}, 1772--1775. National Maritime Museum, Greenwich. \citep{RMG_K1}. Photo: Authors.}
\end{minipage}\hfill
\begin{minipage}[t]{0.48\textwidth}\centering
\includegraphics[height=2.45in,keepaspectratio]{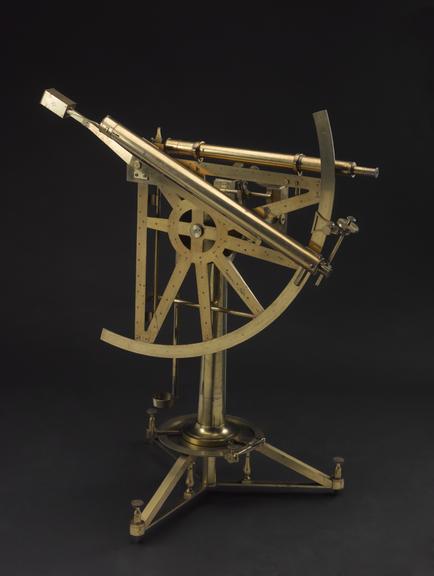}
\caption{\textbf{(b)} Quadrant associated with surveys in Lower Canada, ca. 1780--1820; attributed to Peter and John Dollond. \citep{SMG_Quadrant}. Photo: Board of Trustees, Science Museum, London.}
\end{minipage}
\vspace{0.22in}
\begin{minipage}[t]{0.48\textwidth}\centering
\includegraphics[height=2.65in,keepaspectratio]{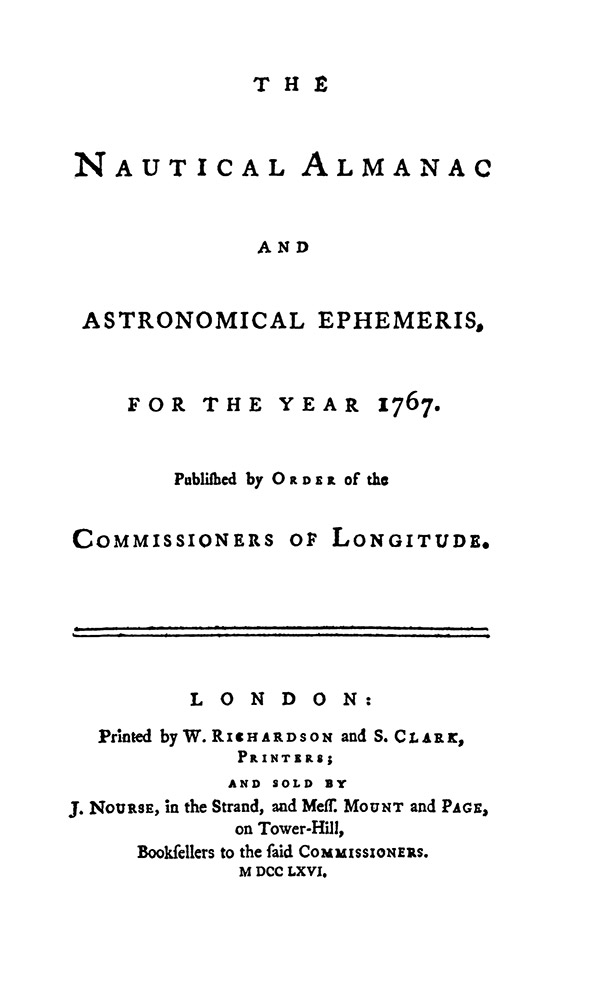}
\caption{\textbf{(c)} Title page of the \emph{Nautical Almanac and Astronomical Ephemeris} for 1767. Smithsonian Institution Libraries. \citep{NauticalAlmanac1767}. Photo: Smithsonian Institution Libraries.}
\end{minipage}\hfill
\begin{minipage}[t]{0.48\textwidth}\centering
\includegraphics[height=2.45in,keepaspectratio]{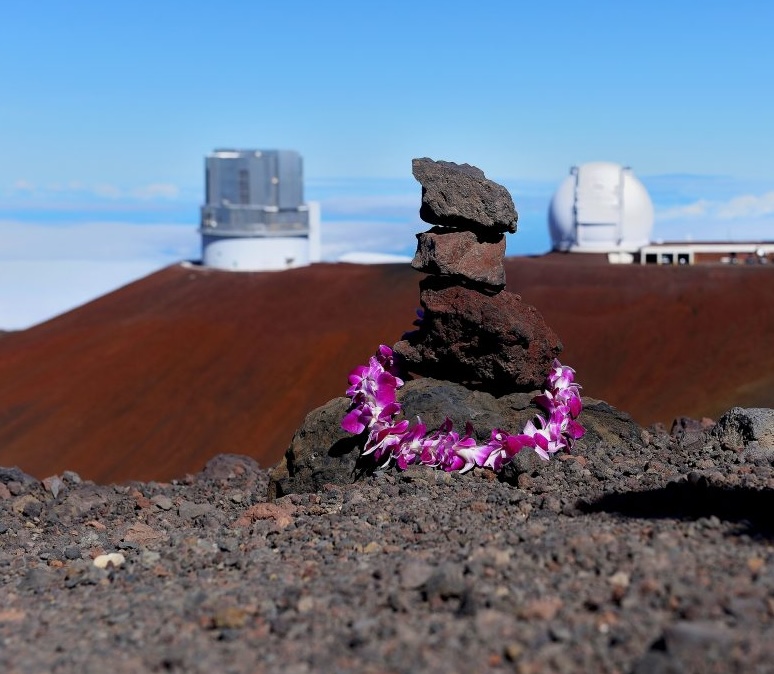}
\caption{\textbf{(d)} Subaru and Keck observatories on Maunakea, 2019. Photo: A. Franceschini/Corbis News.}
\end{minipage}
\end{figure}

\begin{thebibliography}{}

\bibitem[Casumbal-Salazar(2017)]{CasumbalSalazar2017}
Casumbal-Salazar, I. 2017,
A Fictive Kinship: Making ``Modernity,''
``Ancient Hawaiians,'' and the Telescopes on Mauna Kea,
\textit{Native American and Indigenous Studies},
4(2), 1--30.

\bibitem[Commissioners of Longitude(1766)]{NauticalAlmanac1767}
Commissioners of Longitude, 1766,
\textit{The Nautical Almanac and Astronomical Ephemeris
for the Year 1767}
(London: W. Richardson and S. Clark),
Cambridge Digital Library,
\url{https://cudl.lib.cam.ac.uk/view/PR-NAO-01767/7}
(accessed 15 November 2026).

\bibitem[Edney(1997)]{Edney1997}
Edney, M. H. 1997,
\textit{Mapping an Empire: The Geographical
Construction of British India, 1765--1843}
(Chicago: University of Chicago Press).

\bibitem[Forsythe \& Markides(2026)]{ForsytheMarkides2026}
Forsythe, L., \& Markides, J. (eds.) 2026,
\textit{Métis Methodologies}
(Fernwood Publishing).

\bibitem[Kame'eleihiwa(1992)]{Kameeleihiwa1992}
Kame'eleihiwa, L. 1992,
\textit{Native Land and Foreign Desires:
Pehea L\=a E Pono Ai?}
(Honolulu: Bishop Museum Press).

\bibitem[Royal Museums Greenwich(n.d.)]{RMG_K1}
Royal Museums Greenwich, n.d.,
\textit{K1 (Marine Timekeeper)},
National Maritime Museum, Greenwich,
Object ZAA0038,
Collections Online,
\url{https://www.rmg.co.uk/collections/objects/rmgc-object-79143}
(accessed 15 November 2026).

\bibitem[Sadler(1976)]{Sadler1976}
Sadler, D. H. 1976,
Lunar Distances and the Nautical Almanac,
\textit{Vistas in Astronomy}, 20, 113--121.


\bibitem[Science Museum Group(n.d.)]{SMG_Quadrant}
Science Museum Group, n.d.,
\textit{Brass quadrant by Peter and John Dollond,
18 inch radius with optical sights on tripod},
Science Museum Group Collection Online,
Object No. 1911-214,
\url{https://collection.sciencemuseumgroup.org.uk/objects/co56421/brass-quadrant-by-peter-and-john-dollond-18-inch-radius-with-optical-sights-on-tripod}
(accessed 15 November 2026).

\bibitem[Smith(2012)]{Smith2012}
Smith, L. T. 2012,
\textit{Decolonizing Methodologies:
Research and Indigenous Peoples},
2nd ed. (London: Zed Books).

\bibitem[Sorrenson(1996)]{Sorrenson1996}
Sorrenson, R. 1996,
The Ship as a Scientific Instrument
in the Eighteenth Century,
\textit{Osiris}, 11, 221--236.

\bibitem[Tanigawa-Lau(2022)]{TanigawaLau2022}
Tanigawa-Lau, M. 2022,
The State's Kuleana: Deconstructing the Permitting
Process for the Thirty-Meter Telescope and Finding
Restoration Through Systemic Validation of Native
Hawaiian Rights,
\textit{UCLA Law Review}, 68(5), 1390.

\bibitem[Wilson(2008)]{Wilson2008}
Wilson, S. 2008,
\textit{Research Is Ceremony:
Indigenous Research Methods}
(Halifax: Fernwood Publishing).

\end{thebibliography}
\end{document}